\documentclass{ws-procs975x65}

\begin{document}

\title{PAIRING CORRELATIONS IN HALO NUCLEI}

\author{H. SAGAWA}

\address{Center for Mathematical Sciences,    University of Aizu,\\
Ikki-machi, Aizu-Wakamatsu, Fukushima 965-8580, Japan \\
}

\author{K.HAGINO}

\address{Department of Physics, Tohoku University,
 Sendai, 980-8578,  Japan\\
}

\begin{abstract}
Paring correlations in weakly bound halo nuclei $^{6}$He and $^{11}$Li
are  studied by using 
a three-body model with a density-dependent contact interaction.
We investigate  the spatial structure of two-neutron wave function 
in a Borromean nucleus $^{11}$Li.  
The behavior of the neutron pair at different 
densities is simulated by calculating the two-neutron 
wave function at several 
distances between the core nucleus $^9$Li 
and the center of mass of the two
neutrons.  With this representation, 
a strong concentration of the neutron pair 
on the nuclear surface is  quantitatively 
established  for neutron-rich nuclei. 
Dipole excitations in $^{6}$He and $^{11}$Li
   are  also studied within the same three-body 
model and compared with experimental data. The 
 small open  angles between the two neutrons  from the
core are 
extracted empirically by the B(E1) sum rule together with 
the rms mass radii, 
indicating 
the strong di-neutron correlation 
 in the halo nuclei.  
\end{abstract}


\bodymatter
\section{Introduction}

Pairing correlations play a crucial role in many 
Fermion systems, such as liquid $^{3}$He, atomic nuclei, and
ultracold atomic gases. When an attractive
interaction between two Fermions is weak, the pairing correlations  can 
be understood in terms of the well-known 
BCS mechanism, that shows a strong 
correlation in the momentum space. If the interaction is sufficiently
strong, on the other hand, one expects that two Fermions form a 
Bosonic bound state and condense in the ground state of many-body 
system 
\cite{E69}. The transition from the BCS-type pairing
correlation to the Bose-Einstein condensation (BEC) takes place
continuously as a function of the strength of attractive interaction. 
This feature is 
referred to as the BCS-BEC crossover. 

It has been feasible by now to study the structure of nuclei on the edge of
  neutron drip line. Such nuclei are characterized
  by a dilute neutron density
around the nuclear surface so that  one can investigate the
pairing correlations at several densities\cite{MMS05}, 
ranging from the normal density in the center of nucleus to a diluted  
density at the surface.   The pairing correlations 
  are  predicted to be  strong at the
 surface, but  rather weak  at  the normal density and also  far 
outside of the  core.  Thus, 
the weakly bound  nuclei will  provide
 an ideal environment to study the dynamics  
 of pairing correlations in relation with  the BCS-BEC crossover 
  phenomenon.

In this talk, we discuss 
the manifestation of the BCS-BEC crossover phenomenon 
in {\it finite} neutron-rich nuclei. 
We particularly study the ground state wave function 
of a two-neutron halo nuclei, $^{6}$He and  $^{11}$Li. 
These  nuclei  are known to be well described as 
   a three-body system 
consisting of two valence neutrons and the core nucleus($^{4}$He or 
 $^{9}$Li) 
\cite{BF1,BF2,HS05}. 
A strong di-neutron 
correlation as a consequence of pairing interaction between the 
valence neutrons has been shown theoretically 
in $^{11}$Li \cite{BF3,HS05}.  We take 
    this nucleus to 
   study BCS-BEC crossover features in connection to 
the strong two-neutron correlation, which has recently 
been observed  experimentally in low-lying dipole strength in
$^{11}$Li \cite{N06}.

\section{Three-body model and di-neutron correlation}

In order to study the pair wave function in 
$^{11}$Li,  we solve the following  three-body Hamiltonian 
\cite{BF3,HS05}, 
\begin{equation}
H=\hat{h}_{nC}(1)+\hat{h}_{nC}(2)+V_{nn}+\frac{\vec{p}_1\cdot\vec{p}_2}{A_cm}, 
\label{3bh}
\end{equation}
where $m$ and $A_c$ are the nucleon mass and the mass number of the 
inert core nucleus, respectively. 
$\hat{h}_{nC}$ is the single-particle Hamiltonian for a valence 
neutron interacting with the core. 
We use a Woods-Saxon potential 
 for the interaction in $\hat{h}_{nC}$.
The diagonal component of the recoil kinetic energy of the core
nucleus is included in 
$\hat{h}_{nC}$, whereas the off-diagonal part is taken into account
in the last term in the Hamiltonian (\ref{3bh}). 
The interaction between the valence neutrons $V_{nn}$ is 
taken as  a delta interaction whose strength depends
on the density of the core nucleus. Assuming that the core density 
is described by a Fermi function, it reads 
\begin{equation}
V_{nn}(\vec{r}_1,\vec{r}_2)=\delta(\vec{r}_1-\vec{r}_2)
\left(v_0+\frac{v_\rho}{1+\exp[(R-R_\rho)/a_\rho]}\right), 
\label{vnn}
\end{equation}
where $R=|(\vec{r}_1+\vec{r}_2)/2|$. 
We use the same value for the parameters as in 
Refs. \cite{BF3,HS05}. 

The two-particle wave function $\Psi(\vec{r}_1,\vec{r}_2)$ is obtained by
diagonalizing the  three-body Hamiltonian (\ref{3bh}) with a large
model space which is consistent with the $nn$ interaction,
$V_{nn}$. 
To this end, we expand the wave function 
$\Psi(\vec{r}_1,\vec{r}_2)$ 
with the  eigenfunction 
$\phi_{nljj_z}(\vec{r}_i)$ of the single-particle
Hamiltonian $\hat{h}_{nC}$.  
In the expansion, we explicitly exclude those states which 
are occupied by the core nucleus. 

The ground state wave function 
is  obtained as the state with 
the total angular momentum $J=J_z=0$. We transform it to the
coordinate system with the relative and center of mass (cm) 
motions for the valence
neutrons, $\vec{r}=\vec{r}_1-\vec{r}_2$ and 
$\vec{R}=(\vec{r}_1+\vec{r}_2)/2$
\cite{BK67,HSCS07}. 
%
The wave function is first decomposed into the 
total spin $S$=0 and $S$=1 components. Then, 
the coordinate transformation is 
performed  for the $S$=0 component, which is relevant to the
pairing correlation: 
%
\begin{equation}
\Psi^{S=0}(\vec{r}_1,\vec{r}_2)=
\sum_Lf_L(r,R)\,[Y_L(\hat{\vec{r}})Y_L(\hat{\vec{R}})]^{(00)}\,
|\chi_{S=0}\rangle,
\label{tbw}
\end{equation}
where $|\chi_{S=0}\rangle$ is the spin wave function. 
The two-particle wave function is plotted for  $^{11}$Li in two different
coordinates in Fig. \ref{fig1}.   Fig. \ref{fig1}(a) is plotted as a function of 
$r\equiv r_1=r_2$ and the angle between the valence neutrons, while the radial 
coordinates $r$ and $R$ are adopted for Fig. \ref{fig1}(c). The $S=0$ component is only taken for  Fig. \ref{fig1}(b).
One  observes two peaks in both the figures.  The peak at smaller angle 
$\theta_{12}$ in Fig. \ref{fig1}(a) is referred
 to as ``di-neutron configuration'' , while the large angle is called 
``cigar-like configuration''.  One can see the di-neutron configuration
 has a long tail as a typical feature of halo wave function.
%
\def\figsubcap#1{\par\noindent\centering\footnotesize(#1)}
\begin{figure}[b]%
\begin{center}
 \parbox{2.4in}{\includegraphics[scale=1.0,clip]{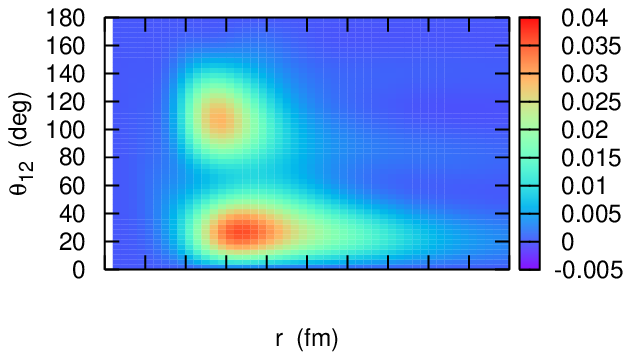}
 \figsubcap{a}}
\hspace*{0.3cm}
\vspace*{-1.0cm}
 \parbox{2.4in}{\includegraphics[scale=1.0,clip]{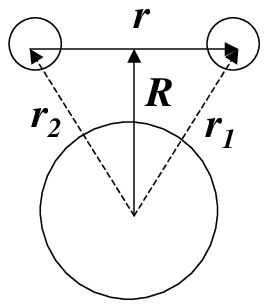}
}
 \parbox{2.4in}{\includegraphics[scale=1.0,clip]{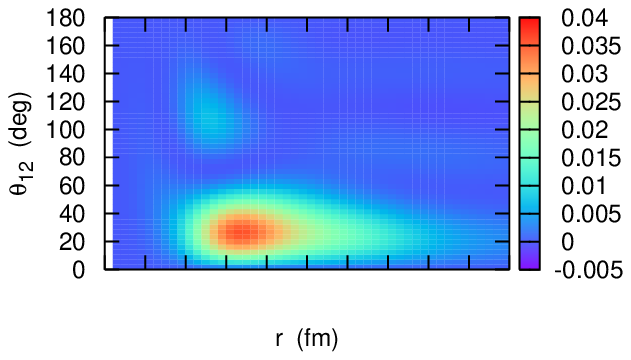}
 \figsubcap{b}}
 \parbox{2.4in}{\includegraphics[scale=1.0,clip]{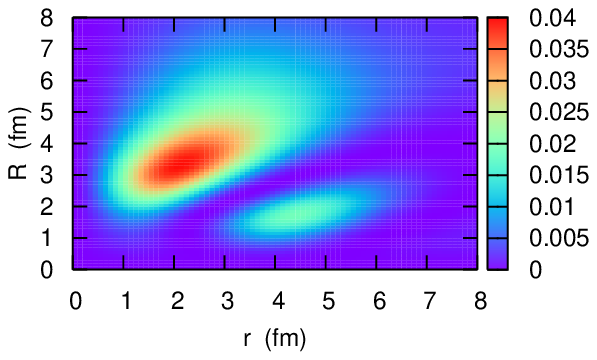}
 \figsubcap{c}}
\caption{A two dimensional  plot of 
 the square of
the  ground state 
two-particle wave function  for
 $^{11}$Li; (a) the total density,  (b) the $S=0$ component
 as a function of the radial coordinate $r\equiv r_1=r_2$
 and the angle between the valence neutrons $\theta_{12}$. (c)
 $r^2R^2|f_{L=0}(r,R)|^2$ in Eq. (\ref{tbw}), 
  as a function of the relative distance $r$ and
the center of mass coordinate $R$ for the valence neutrons as denoted 
in the upper right inset.}
\label{fig1}
\end{center}
\end{figure}
%
Figure \ref{fig1} (c) shows the square of two-particle wave
  function for the $L=0$ 
component.  
One can clearly recognize the two peaked structure 
in the plot, corresponding to 
the di-neutron and the cigar-like configurations.  

\begin{figure}[t]
\begin{center}
\includegraphics[scale=0.4,clip]{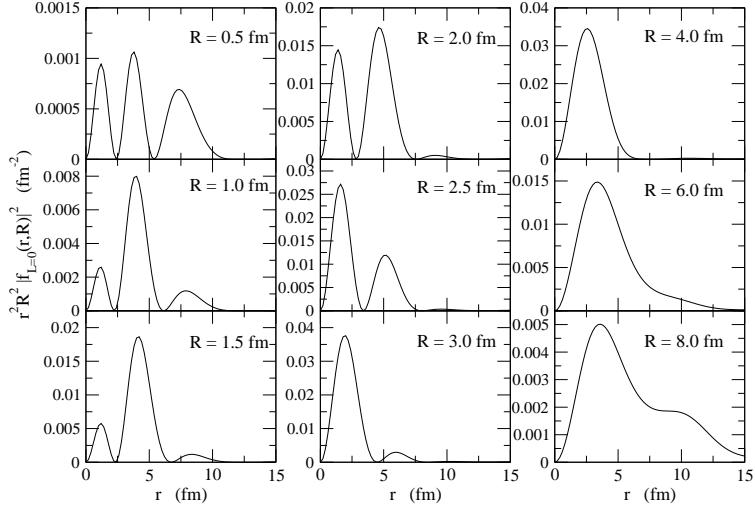}\\
\end{center}
\caption{The  ground state 
two-particle wave functions, $r^2R^2|f_{L=0}(r,R)|^2$ of $^{11}$Li
 as a function of 
the relative distance between 
the neutrons, $r$, at several center of mass distances $R$ as
indicated in the inset.
Notice the different scales on the 
ordinate in the various panels.}
\label{fig2}
\end{figure}

The $L=0$ wave functions of $^{11}$Li 
 for different values of $R$ are  
plotted in Fig. \ref{fig2}.  
Since we consider the density-dependent contact interaction,
Eq. (\ref{vnn}), this is effectively equivalent to probing the wave
function at different densities.  
At $R=0.5$ fm, where the density 
is close to the normal density $\rho_0$, the two particle wave
function is spatially extended and oscillates inside the nuclear
interior. This oscillatory behavior is typical for a Cooper pair
wave function in the BCS approximation, and has in fact been found in 
nuclear and neutron matters at normal density $\rho_0$ \cite{M06,BLS95}.  
As $R$ increases, the density $\rho$ decreases. 
The two-particle wave function then gradually deviates from the BCS-like 
behavior. 
At $R=3$ fm, the oscillatory behavior almost disappears and the 
wave function is largely concentrated inside the first node at 
$r\sim$ 4.5 fm. 
The wave function is compact in shape, indicating the strong 
di-neutron
correlation, typical for BEC 
where many such pairs are present. 
At $R$ larger than 3 fm, 
the squared wave function has 
essentially only one node, and the width of the peak gradually increases as a
function of $R$. 
This behavior is  qualitatively similar  
to  
the pair wave function in infinite 
matter \cite{M06}. We have confirmed using the same three-body model 
that this scenario also holds for another 
Borromean nucleus $^6$He  
as well as for non-Borromean neutron-rich 
nuclei $^{16}$C and  $^{24}$O. 

\begin{figure}[t]
\begin{center}
\includegraphics[scale=0.5,clip]{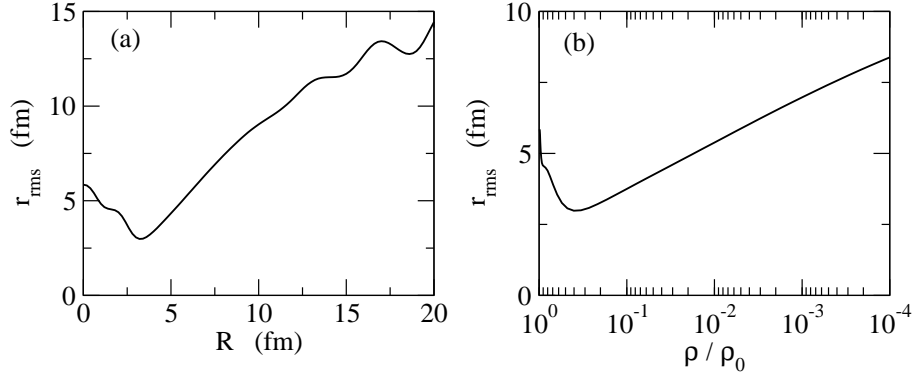}\\
\end{center}
\caption{
The root mean square distance $r_{\rm rms}$ for the neutron pair.  
}
\label{fig3}
\end{figure}

The transition from the BCS-type pairing to the BEC-type di-neutron 
correlation can also clearly be seen in the root mean square (rms)
distance of
the two neutron system. 
We plot this quantity in Fig. 3(a) as a function of $R$. 
In order to compare it with the rms distance in nuclear matter, we
relate the cm 
distance $R$ with the density $\rho$ using
the Fermi-type functional form 
%
$\rho(R)/\rho_0
=[1+\exp((R-R_\rho)/a_\rho)]^{-1}$, 
%
as  used in the $nn$ interaction in Eq. (\ref{vnn}).
Fig. 3(b) shows the rms distance as a function of density $\rho$ thus 
obtained. The rms distance shows a distinct minimum at $\rho\sim 0.4\rho_0$
($R\sim$ 3.2 fm). 
This indicates that the strong di-neutron
correlation grows  in 
$^{11}$Li around this density. 
Notice that the probability to find the two-neutron pair is maximal 
around this region (see Fig. \ref{fig1} (c)). 
The behavior 
 of  rms distance 
as a function of density $\rho$ 
qualitatively well agrees with that in infinite matter 
(see Fig. 3 in Ref. \cite{M06}), although the absolute value of the
rms distance is much smaller in the finite nucleus.  

\section{Dipole excitations and correlation angles}

\begin{figure}[t]
\begin{center}
\includegraphics[scale=0.5,clip]{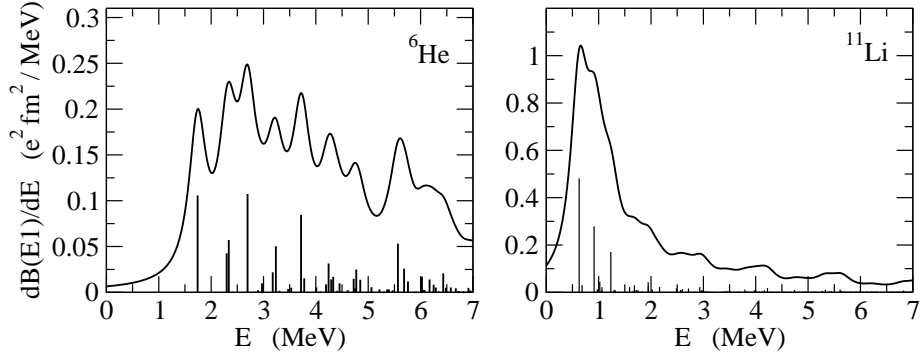}\\
\end{center}
\caption{The B(E1) distribution for the $^6$He and $^{11}$Li nuclei. 
The solid curve is obtained by  a smearing procedure with  a Lorentzian 
weighting factor of  the width 
$\Gamma$=0.2 MeV.}
\label{fig-BE1}
\end{figure}

The rms distance $\sqrt{\langle r^2_{c-2n}\rangle}$ has an intimate
relation to  the  B(E1) strength as
 \cite{BF1,BF2,EHMS07}, 
\begin{equation}
B(E1)=\frac{3}{\pi}\left(\frac{Ze}{A}\right)^2\,
\langle r^2_{c-2n}\rangle. 
\end{equation}
This relation is obtained with closure, which includes unphysical
Pauli forbidden transitions to the states with negative excitation energies.  
Although the effect of Pauli forbidden
transitions is not large, it leads to a non-negligible correction. 
%
In Ref. \cite{EHMS07,HS07a}, it has been proposed to estimate the
experimental value for $\langle r^2_{c-2n}\rangle$ using the relation,
\begin{equation}
\langle r^2_{c-2n}\rangle_{\rm exp} 
= \frac{B(E1; E\leq E_{\rm max})_{\rm exp}}
{B(E1; E\leq E_{\rm max})_{\rm cal}}\,
\cdot\langle r^2_{c-2n}\rangle_{\rm cal}. 
\label{r-c2n}
\end{equation}
The dipole strength distributions for the $^6$He and $^{11}$Li nuclei 
obtained with the three  model are shown in Fig. \ref{fig-BE1}. 
 Also shown by the solid
curves are the B(E1)
distributions smeared with the Lorenzian function with the width of
$\Gamma=0.2$ MeV. For the $^6$He nucleus, we obtain the total B(E1) strength of 0.660
e$^2$fm$^2$ up to $E\leq 5$ MeV and 1.053 
e$^2$fm$^2$ up to $E\leq 10$ MeV.
These are in good agreement with the 
experimental values, 
B(E1; $E\leq$ 5 MeV)=0.59 $\pm$ 0.12 e$^2$fm$^2$ and 
B(E1; $E\leq$ 10 MeV)=1.2 $\pm$ 0.2 e$^2$fm$^2$ \cite{A99}. 
For the $^{11}$Li nucleus, we obtain the total B(E1) strength of 1.405 
e$^2$fm$^2$ up to $E_{\rm rel}=E-S_{2n} \leq 3$ MeV, which is 
compared to the experimental value, 
B(E1; $E_{\rm rel}\leq$ 3 MeV)=1.42 $\pm$ 0.18 e$^2$fm$^2$
\cite{N06}. Again, the experimental data is well reproduced within the 
present model. 
From the calculated values for 
$\langle r^2_{c-2n}\rangle_{\rm cal}$, that is, 13.2 and 26.3 fm$^2$
for $^6$He and $^{11}$Li, respectively, we thus obtain 
$\sqrt{\langle r^2_{c-2n}\rangle_{\rm exp}}=3.878 \pm 0.324$ fm and 
5.15 $\pm$ 0.327 fm for $^6$He and $^{11}$Li, respectively. 
Notice that the value for the $^6$He nucleus is somewhat larger than
the one estimated in Ref. \cite{A99}, that is, 3.36 $\pm$ 0.39 fm. 

\def\figsubcap#1{\par\noindent\centering\footnotesize(#1)}
\begin{figure}[b]
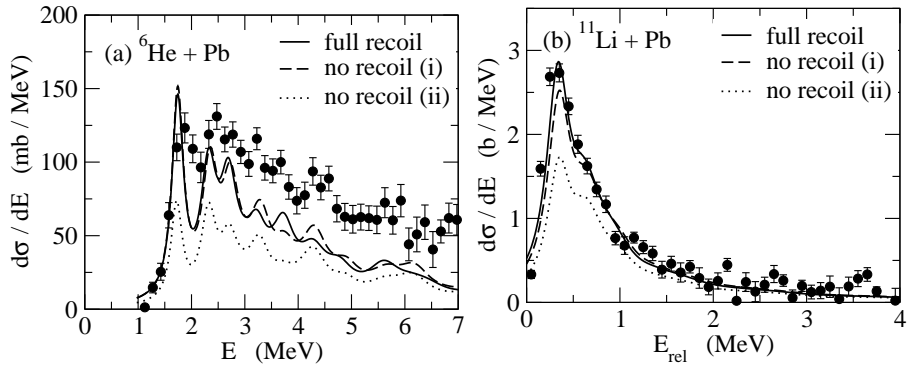
%
\begin{center}
 \parbox{2.4in}{\includegraphics[scale=0.46,clip]{ispun-fig2a}
}
 \parbox{2.4in}{\includegraphics[scale=0.46,clip]{ispun-fig2b}
}
 \caption{Coulomb breakup cross sections 
 (a) for $^6$He+Pb at 240 MeV /nucleon, and 
 (b)  for $^{11}$Li+Pb at 70 MeV /nucleon.
The solid line is the result of the full three-body calculations,
while the dashed and the dotted lines are obtained by treating the recoil 
term approximately (see Ref. \cite{HS07a} for details).
These results are smeared with an energy dependent width of 
$\Gamma = 0.15 \cdot \sqrt{E_{\rm rel}}$ MeV. 
The experimental data are taken from Ref. \cite{A99} for  $^6$He and
from Ref. \cite{N06} for  $^{11}$Li.}
\label{fig-CB}
\end{center}
\end{figure}

We next evaluate the Coulomb breakup cross sections 
based on the relativistic Coulomb excitation theory 
\cite{BB88}. These are
obtained by multiplying the virtual photon number $N_{\rm E1}(E)$ 
to the B(E1) distribution shown in Fig. \ref{fig-BE1}.
  The solid line in Figs. \ref{fig-CB} (a) and (b)
  shows the Coulomb breakup cross sections thus obtained for
$^6$He+Pb reaction at 240 MeV/nucleon \cite{A99} and 
$^{11}$Li+Pb reaction at 70 MeV/nucleon \cite{N06}, respectively. 
In order to facilitate the comparison with the experimental data, we 
smear the discretized cross sections with the Lorenzian function 
with an energy dependent width, $\Gamma = \alpha\cdot \sqrt{E_{\rm rel}}$. 
We take $\alpha=0.15$MeV$^{1/2}$ and 0.25 MeV$^{1/2}$ for $^6$He and 
$^{11}$Li, respectively. We see that the experimental breakup cross
sections are reproduced remarkably well within the present three-body
model, especially for the $^{11}$Li nucleus. 

Let us now discuss the geometry of the $^6$He and $^{11}$Li nuclei. 
Using the experimental value for $\langle r^2_{c-2n}\rangle$ obtained 
from the B(E1) distribution, one can extract the mean opening angle 
between the valence neutrons once an additional information is
available. 
The mean opening angle can be extracted directly when the rms 
distance between the valence neutrons, 
$\langle r^2_{nn}\rangle$, is available. This quantity is related 
to the matter radius and $\langle r_{c-2n}^2\rangle$ 
in the  three-body model \cite{BF1,BF3,EHMS07}, 
\begin{equation}
\langle r_m^2\rangle = 
\frac{A_c}{A}\,\langle r_m^2\rangle_{A_c}
+\frac{2A_c}{A^2}\,\langle r_{c-2n}^2\rangle 
+\frac{1}{2A}\,\langle r_{nn}^2\rangle, 
\label{eq:rnn}
\end{equation}
where $A_c=A-2$ is the mass number of the core nucleus. 
The matter radii 
$\langle r_m^2\rangle$ can be estimated from interaction cross
sections. Employing the Glauber theory in the optical limit, Tanihata
{\it et al.} have obtained 
$\sqrt{\langle r_m^2\rangle}$ = 1.57 $\pm$ 0.04, 
2.48 $\pm$ 0.03, 2.32 $\pm$ 0.02, and 
3.12 $\pm$ 0.16 fm for $^4$He, $^6$He, $^9$Li, and 
$^{11}$Li, respectively \cite{T88}. Using these values, we obtain the 
rms neutron-neutron distance of 
$\sqrt{\langle r_{nn}^2\rangle}$ = 3.75 $\pm$ 0.93 and 
5.50 $\pm$ 2.24 fm for $^6$He and $^{11}$Li, respectively. 
Combining these values with the rms core-di-neutron distance, 
$\sqrt{\langle r_{c-2n}^2\rangle}$, obtained with Eq. (\ref{r-c2n}),
we obtain the mean opening angle of $\langle\theta_{nn}\rangle$ =
51.56$^{+11.2}_{-12.4}$ and 56.2$^{+17.8}_{-21.3}$ degrees for 
$^6$He and $^{11}$Li, respectively. 
These values are close to 
the result of the three-body model calculation, 
$\langle\theta_{nn}\rangle$=66.33 and 65.29 degree for $^6$He and 
$^{11}$Li, respectively \cite{HS05}, although 
the experimental values are somewhat smaller.
An alternative way to 
 extract the value  $\sqrt{\langle r^2_{nn}\rangle}$ was reported
by the three-body 
correlation study in the dissociation of two neutrons in halo nuclei 
\cite{M00,BH07}.  The two neutron correlation function provides the experimental
values for $\sqrt{\langle r_{nn}^2\rangle}$ to be  5.9 $\pm$ 1.2 and 6.6 $\pm$ 1.5 fm for $^6$He, $^{11}$Li, respectively \cite{M00}.
When one adopts the presently obtained value
 for $\sqrt{\langle r_{c-2n}^2\rangle}$ 
with Eq. (\ref{r-c2n}) instead of those in Refs. \cite{N06,A99}, 
one  obtains  
$\langle\theta_{nn}\rangle$=74.5 $^{+11.2}_{-13.1}$ and 
65.2 $^{+11.4}_{-13.0}$  
for $^6$He and $^{11}$Li, respectively. Notice that these values are
in 
good 
agreement with the results of the three-body calculation
\cite{HS05}as is seen  in Table 1. 

\begin{table}  
\tbl{
The geometry of the $^6$He and $^{11}$Li nuclei 
extracted from various experimental data. 
The mean opening angles calculated by the three-body model  
are also given
 in the last line for each nucleus in the table. }
{\begin{tabular}{c|c|cc|c}
\hline
\hline
nucleus & $\sqrt{\langle r_{c-2n}^2\rangle}$ (fm) & 
$\sqrt{\langle r_{nn}^2\rangle}$ (fm)  & method & 
$\langle\theta_{nn}\rangle$ (deg.)  \\
\hline
$^6$He & 3.88$\pm$0.32 & 
3.75 $\pm$ 0.93 & (matter radii) & 51.56 $^{+11.21}_{-12.37}$ \\
&  & 
5.9$\pm$ 1.2 & (neutron correlations) & 74.5 $^{+11.2}_{-13.1}$ \\
   & &      &  &66.33 \cite{HS05}\\
\hline
$^{11}$Li & 5.15$\pm$0.33 & 
5.50 $\pm$ 2.24 & (matter radii) & 56.2 $^{+17.8}_{-21.3}$ \\
&  & 
6.6$\pm$ 1.5 & (neutron correlations) & 65.2 $^{+11.4}_{-13.0}$ \\
   & &      &   & 65.29 \cite{HS05} \\\hline
\hline
\end{tabular}}
\end{table}

\section{Summary}
We studied the two-neutron (2n) wave function 
in the Borromean nuclei $^{6}$He and 
 $^{11}$Li by using the three-body model with 
the density-dependent pairing force.  
We explored the spatial distributions of 
2n 
wave function as a function of 
the cm 
distance $R$ from the core nucleus and  
   found that the  structure
 of the 
2n 
wave function alters drastically as $R$
 is varied.  
We  also showed that the relative distance  
between  the 
two neutrons  
scales consistently to that in the infinite 
matter as a function of density. 
These features are
  in close analogue to the characteristics of 
the BCS-BEC crossover phenomenon found in the infinite nuclear and neutron
matters.
We have used the same  three-body model 
to analyze the B(E1) distribution as well as the
Coulomb breakup cross section of the $^6$He and $^{11}$Li nuclei. 
We have shown that the strong concentration of the B(E1) strength near 
the continuum threshold can be well reproduced with the present model
for both the nuclei.
Using the calculated B(E1) strength, we extracted 
the experimental value for the rms distance between the 
core and di-neutron, which was then converted to the mean opening angle
of the two valence neutrons. 
We have found that the mean opening angles thus obtained are in good
agreement of the results of the three-body model calculation. 

\section*{Acknowledgements}
  
We thank H. Esbensen, J. Carbonell, and P. Schuck for fruitful 
collaborations  which made this presentation possible.
  We thank also T. Aumann and K. Nakamura
 for valuable experimental information and discussions.


\begin{thebibliography}{99}




\bibitem{E69}D.M. Eagles, Phys. Rev. {\bf 186}, 456 (1969); 
A.J. Leggett, J. Phys. C{\bf 41}, 7 (1980); 
P. Nozi\`eres and S. Schmitt-Rink, 
J. Lowe Temp. Phys. {\bf 59}, 195 (1985). 



\bibitem{MMS05}M. Matsuo, K. Mizuyama, and Y. Serizawa,
  Phys. Rev. C{\bf 71}, 064326 (2005). 

\bibitem{BF1}
G.F.  Bertsch and  H. Esbensen, Ann. Phys. (N.Y.) {\bf 209}, 327
(1991).

\bibitem{BF2}  H. Esbensen  and  G.F. Bertsch, Nucl. Phys. {\bf A542}, 310
  (1992).

\bibitem{HS05}K. Hagino and H. Sagawa, Phys. Rev. C{\bf 72}, 
044321 (2005). 



\bibitem{BF3} H. Esbensen, G.F. Bertsch and K. Hencken,
  Phys. Rev. C{\bf 56}, 3054 (1999).

\bibitem{N06}T. Nakamura {\it et al.}, Phys. Rev. Lett. 
{\bf 96}, 252502 (2006). 




\bibitem{BK67}B.F. Bayman and A. Kallio, Phys. Rev. {\bf 156}, 1121
  (1967). 


\bibitem{HSCS07}K. Hagino, H. Sagawa, J. Carbonell, and P. Schuck, 
Phys. Rev. Lett. {\bf 99}, 022506 (2007). 

\bibitem{M06}M. Matsuo, Phys. Rev. C{\bf 73}, 044309 (2006). 
\bibitem{BLS95}M. Baldo, U. Lombardo, and P. Schuck, 
Phys. Rev. C{\bf 52}, 975 (1995).


\bibitem{EHMS07}H. Esbensen, K. Hagino, P. Mueller, and 
H. Sagawa, Phys. Rev. C{\bf 76}, 024302 (2007). 

 
\bibitem{HS07a}  K. Hagino and H. Sagawa, Phys. Rev. C, in press (2007).

\bibitem{BB88}C.A. Bertulani and G. Baur, Phys. Rep. {\bf 163}, 299
  (1988). 



\bibitem{A99}T. Aumann {\it et al.}, Phys. Rev. C{\bf 59}, 1252 (1999). 


\bibitem{T88}I. Tanihata {\it et al.}, Phys. Lett. B{\bf 206}, 592
  (1988); A. Ozawa {\it et al.}, Nucl. Phys. {\bf A693}, 32 (2001). 

\bibitem{M00}F.M. Marques {\it et al.}, Phys. Lett. B{\bf 476}, 219
  (2000).

\bibitem{BH07}C.A. Bertulani and M.S. Hussein, 
arXiv:0705.3998. 

\end{thebibliography}
\end{document}